\documentclass[twocolumn,showpacs,amsmath,amssymb]{revtex4}

\usepackage{amsmath}
\usepackage{graphicx}
\usepackage{amsfonts}
\usepackage{dcolumn}
\usepackage{bm}
\usepackage{CJK}

\begin{document}

\title{Flexibility Induced Motion Transition of Active Filament: Rotation without Long-range Hydrodynamic Interaction}

\author{Huijun Jiang} \email{hjjiang@ustc.edu}
\author{Zhonghuai Hou} \email{hzhlj@ustc.edu.cn}

\affiliation{Department of Chemical Physics \& Hefei National Laboratory for Physical Sciences at
 the Microscale, University of Science and Technology of China,Hefei, Anhui 230026, China}

\date{\today}

\begin{abstract}
We investigate the motion of active semiflexible filament with shape kinematics and hydrodynamic interaction including. Three types of filament motion are found: Translation, snaking and rotation. Change of flexibility will induce instability of shape kinematics and further result in asymmetry of shape kinematics respect to the motion of mass center, which are responsible to a continuous-like transition from translation to snaking and a first-order-like transition from snaking to rotation, respectively. Of particular interest, we find that long-range hydrodynamic interaction is not necessary for filament rotation, but can enhance remarkably the parameter region for its appearance. This finding may provide an evidence that the experimentally found collective rotation of active filaments is more likely to arise from the individual property even without the long-range hydrodynamic interaction.
\end{abstract}
\maketitle

Due to its important role for cell division, cell motility, force generation\cite{Bray2001,Howard2001} and self-assembly of active soft matter\cite{SoM11003050}, motion of filament at subcellular level has gained rapidly increasing attentions\cite{Nat10000073,Nat12000448,PRL06258103,PRL12258101,SMa11003213}. Usually, the dynamics of filament is governed by low Reynolds number hydrodynamics, where viscous effects dominate and inertial effects are negligible\cite{happel1965}, and involves the active process by motor proteins walking on it, which convert chemical energy into work and generate propulsions. So far, several theoretical models have been proposed to describe the active filament. For example, a rod model is used to study enhanced ordering of interacting filament by molecular motors\cite{PRL06258103}, and a structure consists of a spherical head, a curved midpiece and a beating tail has been constructed to reveal the hydrodynamics of sperm cells near surfaces\cite{BPJ10001018}. Lately, a lattice-based cellular automaton model is implemented to reproduce wave-like collective motion patterns\cite{Nat10000073}. Nevertheless, these models either neglect the shape kinematics of each filament part evolving dynamically as a result of interaction between the actuation, material elasticity, and viscous drag\cite{PRL12258101}, or miss the hydrodynamics induced long-range interaction\cite{Nat10000073}, both of which are important in understanding the active motion of filament. Recent experiments in planar geometry motility assay reveals that active filaments will self-organize to form coherently moving structures, such as propagating density waves and vortex motion\cite{Nat10000073,Nat12000448}. While Schaler \textit{et al} show that long-range hydrodynamic interactions play a crucial role in the pattern forming mechanisms\cite{SMa11003213}, Sumino and his collaborators believe that the emergence of vortex structures is the result of the instinct rotation motion of single filament in combination with local interactions and is no need for long-range interactions\cite{Nat12000448}. This argument then calls for a complete study of motion of single active filament.

In this Letter, we investigate the motion of active filament based on a simple theoretical model including both long-range hydrodynamic interaction (HI) and shape kinematics. We find that flexibility is crucial for the motion of filament by controlling the shape kinematics. For complete rigidity, activation results in a translation movement. By a qualitative analysis and simulations, we show that, as the rigidity $\kappa$ decreases, filament undergoes a transition from translation to snaking at $\kappa=\kappa_1$ due to a flexibility induced instability of shape kinematics. By further decreasing rigidity, an interesting rotation of filament can be observed for $\kappa\leq\kappa_2$ as a result of asymmetry of shape kinematics respect to the motion of mass center. The scaling behavior of the continuous-like transition near $\kappa_1$ is revealed, and mixed-type movement of filament is found near $\kappa_2$ indicating a first-order-like snaking-rotation transition. Particular attention is paid on the effect of long-range HI on the rotation of filament. The comparison between filament motion with and without long-range HI shows that, long-range HI is not a necessary component for the filament rotation, but can amplify it boundary $\kappa_2$ considerably, which indicates that more rigid filament can also rotate when long-range HI is included.

\begin{figure}
\begin{center}
\includegraphics[width=0.8\columnwidth] {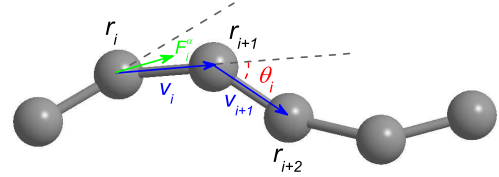}
\end{center}
\caption{(a) Schematic diagram of the bead-rod model of the active filament.} \label{fig:model}
\end{figure}

\textit{Model and method}--Our simulation is based on a bead-rod model consists of $N$ beads whose position is ${\bf{r}}_i, i=1,...,N$, and $N-1$ inextensible rods with rod vector ${\bf{v}}_i={\bf{r}}_{i+1}-{\bf{r}}_i, i=1,...,N-1$, whose schematic diagram is shown in Fig.\ref{fig:model}. The motion of filament is given by the following equation
\begin{equation}
\dot{\bf{R}}={\bf{M}}(-\frac{\partial U^f}{\partial {\bf{R}}}-\frac{\partial U^{ev}}{\partial {\bf{R}}}+{\bf{F}}^\alpha)+{\bf{B}}\varsigma(t).
\label{eq:DE}
\end{equation}
where ${\bf{R}}=({\bf{r}}_1,...,{\bf{r}}_N)^T$. The flexibility of filament is realized by a bending potential
\begin{equation}
U^f({\bf{r}}_i,{\bf{r}}_{i+1},{\bf{r}}_{i+2})=\frac{1}{2}\kappa(\frac{\theta_i}{\pi})^2, i=1,...,N-2
\label{eq:Uf},
\end{equation}
where $\theta_i$ is the angle between two adjacent rod vectors ${\bf{v}}_{i+1}$ and ${\bf{v}}_i$. The exclusive volume effect is measured by
\begin{equation}
U^{ev}(r_{ij})=\begin{cases}
    4\xi\{(\frac{2a}{r})^{12}-\frac{2a}{r})^6\}+\xi, &r_{ij}<2\sqrt[6]{2}a\\
    0,&r_{ij}\geq2\sqrt[6]{2}a
\end{cases}
\label{eq:ev},
\end{equation}
in which $r_{ij}$ is the distance between bead $i$ and $j$. The activation on each rod segment is considered to be a constant force $2\alpha$ along the rod vectors, which is coarsen from a detailed rod model\cite{PRL06258103} and neglect the transverse stretch of motors for simplicity. Then, the active force on bead $i$ is the resultant of ones on rod $i-1$ and $i$,
\begin{equation}
{\bf{F}}^\alpha_i=\alpha\begin{cases}
    \hat{\bf{v}}_i, &i=1\\
    \hat{\bf{v}}_i+\hat{\bf{v}}_{i-1},&1<i<N\\
    \hat{\bf{v}}_{i-1},&i=N
\end{cases}
\label{eq:active},
\end{equation}
where $\hat{\bf{v}}_i={\bf{v}}_i/|{\bf{v}}_i|$ is the unit vector of ${\bf{v}}_i$. HI is taken into account by using the Rotne-Prager-Yamakawa approximation. The elements of mobility matrix ${\bf{M}}$ is given by
\begin{equation}
{\bf{m}}_{ij}=\mu\begin{cases}
    {\bf{1}}, {\kern 118pt}i=j\\
    [\frac{3}{4}\frac{a}{r_{ij}}({\bf{1}}+\hat{\bf{r}}_{ij}\otimes\hat{\bf{r}}_{ij})+\frac{1}{2}(\frac{a}{r_{ij}})^3({\bf{1}}-3\hat{\bf{r}}_{ij}\otimes\hat{\bf{r}}_{ij})],\\ {\kern 128pt}r_{ij}\geq2a, i\neq j\\
    [(1-\frac{9r_{ij}}{32a}){\bf{1}}+\frac{3r_{ij}}{32a}\hat{\bf{r}}_{ij}\otimes\hat{\bf{r}}_{ij})], {\kern 3pt}r_{ij}<2a, i\neq j\\
\end{cases}
\label{eq:HI}.
\end{equation}
The diagonal elements present the local friction and others the long-range interaction. The self mobility $\mu=1/(6\pi\eta a)$ of a spherical particle is related to the Stokes friction coefficient, where $a$ is the bead radius, and $\eta$ is the viscosity of the fluid. $\varsigma(t)$ is a set of independent $(0,1)$ Gaussian white noise, and the amplitude matrix obeys ${\bf{B}}{\bf{B}}^T=2k_BT{\bf{M}}$.  In simulation, rod constraints are usually ensured by SHAKE algorithm\cite{SHAKE77000327}. For convenience, we describe rods in our simulation by a FENE-fraenkel spring with very large spring constant\cite{JCP06044911}.
\begin{equation}
{\bf{F}}=\frac{H(|{\bf{v}}|-1)}{1-(1-|{\bf{v}}|)^2/\sigma^2}\frac{{\bf{v}}}{|{\bf{v}}|},  (1-\sigma)<|{\bf{v}}|<(1+\sigma),
\end{equation}
where $H$ is the spring constant and $\sigma$ is the extensibility parameter that defines the maximum possible deviation between the actual spring length and the natural length. To prevent overstretching or overcompressing the springs, an implicit predictor-corrector method is employed to calculate the rod length, which has been described in detail in\cite{JNN03000141}. The length, time and energy scale are chosen as diameter of bead $2a$, integration time step $dt$ and $\xi$. We fix $N=10$, $L=2(N-1)=18$ and $\alpha=4$ if not otherwise stated. All the results are derived by simulations for at least $10^7$ time steps.

\begin{figure}
\begin{center}
\includegraphics[width=1.0\columnwidth] {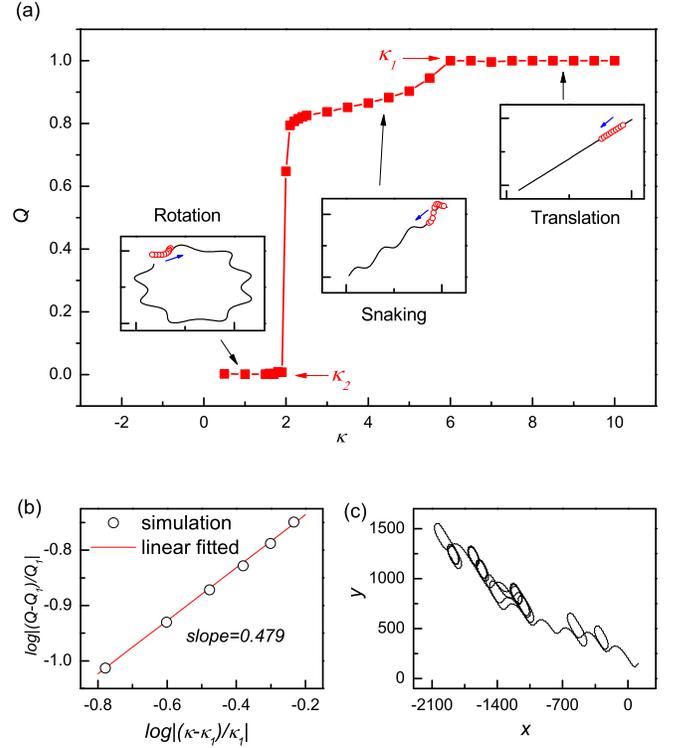}
\end{center}
\caption{Flexibility induced motion transition of active filament. (a) Order paremeter $Q$ (time-average velocity of mass center) as a function of flexibility $\kappa$. Insert panel of (a): Three types of filament motion (presented by trajectory of mass center, solid line), translation ($\kappa=8$), snaking ($\kappa=4$) and rotation ($\kappa=1.5$). (b) Scaling behavior near the translation-snaking transition point $\kappa_1$. The simulation result can be well fitted by a line with slope $0.479$ (c) A mixed-type of active filament motion observed near the snaking-rotation transition point $\kappa_2$.} \label{fig:OP}
\end{figure}

\textit{Result}--Three types of motions (presented by trajectory of mass center of the filament) are observed in different region of filament flexibility. In the right insert panel of Fig.\ref{fig:OP}(a), the filament is of straight line shape, and also moves forward along a straight line (translation). The middle insert panel of Fig.\ref{fig:OP}(a) shows a snaking movement of the filament while the shape of filament is no longer a straight line but becomes cilialike beating. Interesting motion is found when the filament is flexible enough (the left insert panel of Fig.\ref{fig:OP}(a)). The filament rotates spontaneously within the effect of activation, flexibility and HI.

To take a global picture of the filament motion, we define an order parameter to distinct these three types of movements.
\begin{equation}
Q=\frac{1}{C_0}|\lim_{t\rightarrow\infty}\int_0^t \dot{\bf{R}}_{MC}dt|,
\label{eq:OP}
\end{equation}
where ${\bf{R}}_{MC}$ is the coordinate of mass center, and  $C_0$ is value of stationary rate of translation movement for the same active force.  In Fig.\ref{fig:OP}(a), we present a quantitative description of the filament motion as a function of flexibility. It is found that, as $\kappa$ decreases, filament undergoes two motion transitions: the first transition from translation to snaking at $\kappa_1=6.0$, and the second from snaking to rotation at $\kappa_2=1.9$. A scaling behavior near the continuous-like transition point $\kappa_1$ is given in Fig.\ref{fig:OP}(b). The simulated dependence of $log|(Q-Q_1)/Q_1|$ on $log|(\kappa-\kappa_1)/\kappa_1|$ can be well fitted by a line with slope $0.479$, implying that the snaking movement may arises from an instability of translation occuring via a supercritical bifurcation. In Fig.\ref{fig:OP}(c), we record a mixed-type of motion near $\kappa_2$. The filament first snakes for a while, then rotates for a short period and turns back to snaking. This process is repeated over and over again. The coexistence of snaking and rotation indicates that the filament motion is multistable near the transition point $\kappa_2$, which also suggests that the transition from snaking to rotation may be a first-order like transition.

\begin{figure}
\begin{center}
\includegraphics[width=0.7\columnwidth] {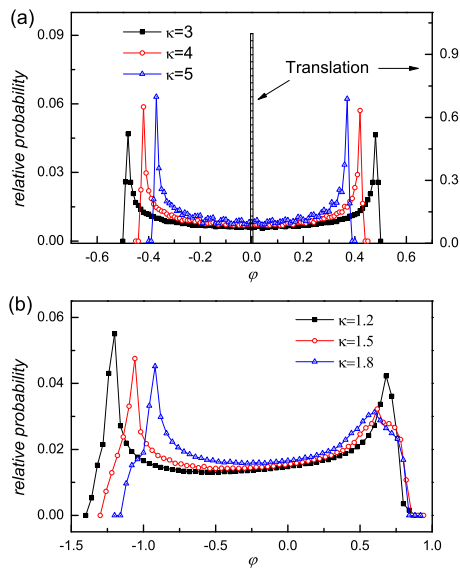}
\end{center}
\caption{The stability and symmetry of shape kinematics respect to the motion of mass center, which is presented by the distribution of angle $\varphi$ between velocity of the first bead and the mass center for the motion of (a) snaking and (b) rotation. The distribution of $\varphi$ for translation is also plotted in (a) in patterned bar.} \label{fig:dis}
\end{figure}

We then do further studies about the underlying mechanism of these two transitions. The result reveals that motion transitions are closely related with the shape kinematics of filament. Firstly, a qualitative linear stability analysis can be given by taking the last 3-bead segment as a sample. When the shape of straight line is stable and neglect long-range HI, the shape kinematics of the 3-bead segment is given by $\dot{\theta}\propto\alpha sin\theta-\kappa\theta/\pi$. The straight line solution $\theta=0$ loses its stability for $\kappa\leq\alpha\pi$. To verify this instability, the distributions of the angle $\varphi$ between velocity of the first bead and of the mass center are shown in Fig.\ref{fig:dis}(a). For translation, $\varphi$ only distributes near  $\varphi=0$. Once filament moves in type of snaking, two symmetric peaks of $\varphi$ emerge, which indicates that straight line shape does lose its stability at $\kappa=\kappa_1$, and results in a shape kinematics of cilialike beating. The distributions for $\kappa=3, 4$ and $5$ show that decreasing of $\kappa$ amplifies the amplitude of cilialike beating. Then, we then simulate the distribution of $\varphi$ for rotation (Fig.\ref{fig:dis}(c)). Similar to the one of snaking, the distribution shows two peaks, too, and the amplitude of cilialike beating is further enlarged. However, these two peaks are no longer symmetric to each other in both positions and the relative probabilities. As a result of the asymmetry of shape kinematics respect to the motion of mass center, the filament keeps tuning clockwise or anticlockwise which is determined by the preference of the distribution. Cumulation of the tuning finally forms the rotation motion of active filament.

\begin{figure}
\begin{center}
\includegraphics[width=0.8\columnwidth] {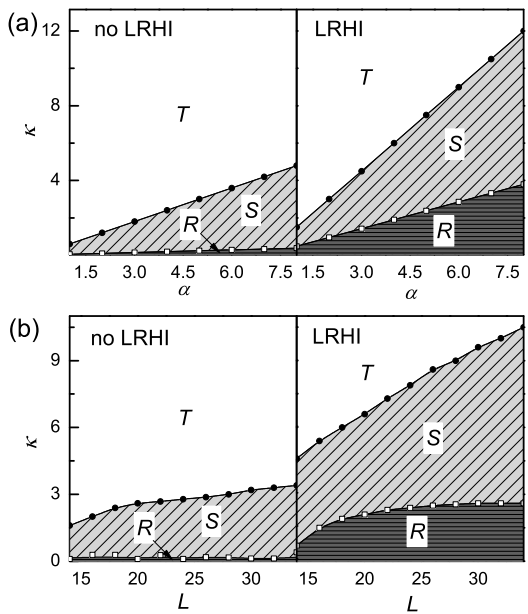}
\end{center}
\caption{Phase diagram of active filament motion on (a) $\alpha-\kappa$ and (b) $L-\kappa$ plane  without (left) and with (right) long-range HI (LRHI). T: Translation. S: Snaking. R: Rotation. The translation-snaking (T-S) and snaking-rotation (S-R) boundary are plotted by line-cycle and line-square respectively. We fix filament length $L=18$ for (a) and active force $\alpha=4$ for (b).} \label{fig:hi}
\end{figure}

Now, we try to reveal the effect of long-range HI on the motion of active filament. In our simulation, long-range HI can be conveniently turned off by considering only the diagonal elements of mobility matrix ${\bf{M}}$. Surprisingly, in contrary to the intuition that long-range HI plays important role in formation of different types of filament motion, all the three types of filament motion are still observed in simulation without long-range HI. The transitions from translation to snaking and from snaking to rotation are similar to their respective one with long-range HI, too. To show this result clearly, we plot the phase diagram of active filament motion with and without long-range HI in $\alpha-\kappa$ plane in Fig.\ref{fig:hi}(a). It can be seen that, no matter long-range HI presents or not, three distinct motion regions can be identified clearly. In other word, long-range HI is not relevant to the formation of active filament motion. Besides of this, several other observation can be concluded from Fig.\ref{fig:hi}(a). Firstly, the instability of straight line shape occurs at $\kappa_1\propto \alpha$, which is accordant with the qualitative linear stability analysis. Determining of the exact instability condition may be mathematically complicated, and needs further studies. Secondly, both the transition boundaries with and without long-range HI are proportional to the active force except that the coefficient are different. This implies that the long-range HI will result in an effective coefficient of the instability. Finally, the filament motion in type of rotation or snaking are enhanced by long-range HI. By a comparison of  flexibility region in which filament moves in type of rotation or snaking with and without long-range HI for different active force $\alpha$, it can be seen that, when long-range HI presents, the range of rotation region $\kappa\leq\kappa_2$ is of magnitude larger than the one only considering local HI. In the meanwhile, width of snaking region $\kappa_1-\kappa_2$ is also enlarged several times by long-range HI. Thus, Fig.\ref{fig:hi}(a) gives evidences that single filament does rotate without long-range HI, and the parameter region for rotation can be enhanced by long-range HI. This finding supports that the shape kinematics is sufficient to form instinct rotation of single active filament, and long-range HI  may be not necessary for the formation of collective vortex of filaments\cite{Nat12000448}.

Besides of long-range HI, the effect of activation $\alpha$ on the motion of filament can also be concluded from Fig.\ref{fig:hi}(a). For a given $\kappa$, $\alpha$ will induce motion transition of filament, too. Similar to decreasing of $\kappa$, the filament first undergoes transition from translation to snaking, then the transition from snaking to rotation, by increasing $\alpha$. It is noted that, when long-range HI absents, the slope of snaking-rotation boundary is very small (Fig.\ref{fig:hi}(a)). As a result, rotation may not be observed by finite increasing of $\alpha$. When long-range HI presents, however, the slope of snaking-rotation boundary becomes much sharper, and rotation will be more easily observed in simulation. At last, the effect of filament length $L$ is studied. The simulated phase diagram in $L-\kappa$ plane in Fig.\ref{fig:hi}(b) shows that, when long-range HI presents, proportionable $\kappa_1$ is needed for the transition from translation to snaking as a result of increasing of $L$. The snaking-rotation transition point $\kappa_2$, in contrary, saturates at about $\kappa_2=2.6$. Compared with ones considering only local HI,  long-range HI can also enhances the rotation and snaking region. In addition, we also simulate the motion of much longer filament. When $\kappa$ is small, the filament rolls into a coil state, and rotation is hardly to be observed in simulations no matter including long-range HI or not. On the other hand, for sufficient large $\kappa$, the translation is still stable and transition from translation to snaking can always present.

\textit{Conclusion and discussion}--Flexibility induced motion transition has been studied by a simple active filament model which allows consideration of both long-range HI and shape kinematics. This research reveals that the motion of filament is mainly determined by flexibility via changing the stability and symmetry of shape kinematics. We also demonstrate that long-range HI can enhance but not crucial for the motion of single filament such as rotation. These findings may be helpful to uncover the underlying mechanism of collective behaviors observed in experiments. What's more,  based on the simple model of single active filament, building up a step by step theoretical framework which relates the individual shape kinematics and long-range HI to the nonequilibrium self-assembly of active filaments is possible.

When we finish this Letter, we notice that similar work on active filament has been done by Jayaraman \textit{et al} and reported in the latest Physical Review Letters\cite{PRL12158302}. They have found that the interplay between long-range hydrodynamics and semiflexibility is necessary for the rotation motion of the filament: A bending instability of an initially straight filament spontaneously breaks flow symmetry and leads to autonomous filament motion which, depending on
conformational symmetry, can be translational or rotational. The difference role of long-range HI revealed in their work and ours may need further understanding.

\textit{Acknowledgments}--This work is supported by National Science Foundation of China(20933006, 91027012).


\end{document}